\documentclass[twocolumn,showpacs,preprintnumbers,amsmath,amssymb]{revtex4}
\usepackage{graphicx}% Include figure files
\usepackage{dcolumn}% Align table columns on decimal point
\usepackage{bm}% bold math
\usepackage{epsfig}
\newcommand{\be}{\begin{eqnarray}}
\newcommand{\ee}{\end{eqnarray}}

\def\be{\begin{equation}}
\def\ee{\end{equation}}
\def\bea{\begin{eqnarray}}
\def\eea{\end{eqnarray}}

\begin{document}
\title{Polaron, molecule and  pairing  in  one-dimensional spin-1/2 Fermi gas with an attractive  Delta-function interaction}

\author{ Xiwen Guan}
\affiliation{Department of
Theoretical Physics, Research School of Physics and Engineering,
Australian National University, Canberra ACT 0200, Australia}

\date{\today}

\begin{abstract}

Using   solutions of the  discrete Bethe ansatz equations,  we study in detail the quantum impurity problem of  a spin-down fermion immersed into a   fully ploarized spin-up Fermi sea with weak attraction. We prove that  this impurity fermion in  the one-dimensional (1D)  fermionic medium behaves like a   polaron   for weak attraction.   However, as the attraction grows,  the spin-down fermion binds with one  spin-up fermion  from the  fully-polarized medium to form a tightly bond molecule. Thus it is seen that the system undergos a cross-over from a mean field polaron-like nature into  a mixture of excess fermions and a bosonic molecule as the attraction changes from  weak attraction  into  strong attraction.  This polaron-molecule cross-over is universal in 1D many-body systems of interacting fermions.  In thermodynamic limit,   we   further study the relationship between  the Fredholm  equations  for the 1D spin-1/2  Fermi gas  with  weakly repulsive and attractive delta-function interactions.   
 \end{abstract}


\maketitle

%\section{The 1D $SU(3)$ Fermi gas}

%\subsection{Repulsion}
\section{Introduction}

The study of one-dimensional (1D)  spin-1/2 Delta-function interacting  Fermi gas \cite{Yang,Gaudin}   is an active area of research in the field of cold atoms. The fundamental physics of the model with arbitrary spin population imbalance are determined by the set of transcendental equations  which were found by Yang \cite{Yang} using the Bethe ansatz hypothesis in 1967. 
The model  displays a  remarkable Fulde-Ferrell-Larkin-Ovchinnikov (FFLO)-like pairing \cite{Fulde1964,Larkin1965,Yang2001,Feiguin2007,Tezuka2008,Zhao2008}, quantum phase transitions and quantum critical phenomena \cite{Erhai,Guan-Ho,Yin}.  It  has a novel phase diagram  caused by a difference in the  number of spin-up and spin-down atoms \cite{Orso,Hu,Guan2007,Mueller,Wadati}.  The key feature of the phase diagram was experimentally confirmed  by Liao et al  \cite{Exp8}  at Rice University  in the strongly attractive regime of  fermionic ${}^6$Li atoms confined  to  the two lowest sub-hyperfine states. 

In 3D, for weak attraction limit,  a spin-down fermion propagates almost freely in a spin-up medium. As attraction increases the spin-down atom is dressed with the localized cloud of scattered surrounding fermions constituting the Fermi polaron \cite{Polaron-1,Polaron-2,Polaron-3}. However, if one considers a small portion of spin-down fermions immersed into a fully polarized spin-up medium with strong attraction, it is seen that the system undergos a  phase transition from a fully-polarized Fermi gas into a mixture of excess fermions and bosonic molecules \cite{Bruun,Mathy}.      
%It is natural  to ask  what  significant nature of  the system possesses   for a single spin-down fermion immersed into a fully-polarized %Fermi sea with  weak and strong  attractions in 1D?   

Using variational ansatz Parish showed  that there does not exist a true polaron-molecule  transition in the 1D highly polarized  Fermionic system  \cite{Parish}.  In fact, this   variational ansatz  is not valid to capture the quasiparticle behaviour for  the 1D interacting Fermi gas because it gives a divergent integral. The polaron-like effect can  persist in the  1D many-body system \cite{Leskinen}. 
In the present paper, using asymptotic solution of the Bethe ansatz equations, we present an  analytical study of the quantum impurity problem in 1D Fermionic medium.  We prove  that a  spin-down fermion immersed into fully polarized spin-up medium with weak attraction is dressed with surrounding fermions and behaves like a Fermi  polaron, also see \cite{note}. The spin-down fermions receive a mean field from the fully-polarized Fermi sea.   In this limit, decoupling the two spin components gives a polaron-like quasiparticle associating with the spin-down fermion dressed by the particle-hole excitations. The mean field binding energy and the effective mass of the polaron can be analytically calculated from the discrete Bethe ansatz equations. However, as an attractive interaction grows, the spin-down fermion binds with one spin-up fermion from the medium to form a   tightly bound molecule. The cross-over is evidenced by the changes from a mean field attractive binding energy of the polaron with an effective mass $m^{*}= m$ to the binding energy of the single molecule with an effective mass $m^{*}= 2m$ as the attraction grows from $c=0$ to $c=-\infty$. Here $m$ is the actual mass of the fermions, and $c$ is the interaction strength.
Thus the system undergos   a cross-over from a mean field polaronic nature into  a mixture of excess fermions and a bosonic molecule as the attraction changes from weak into strong attractions.  Furthermore, we   discuss  the relationship between  the Fredholm  equations  for the 1D spin-1/2  Fermi gas  with  weakly repulsive and attractive Delta-function interactions.

\section{ The model}
\label{model}

The  model Hamiltonian \cite{Yang,Gaudin}
\begin{eqnarray}
{\cal H} &=& \sum _{\sigma=\downarrow,\uparrow} \int \phi _{\sigma}^{\dagger}(x) \left
(-\frac{\hbar^{2}}{2m}\frac{d^{2}}{dx^{2}} + \mu_{\sigma} \right ) \phi
_{\sigma}^{}(x) dx\nonumber\\
&& +  g_{\rm 1D} \int \phi _{\downarrow}^{\dagger}(x) \phi
_{\uparrow}^{\dagger}(x) \phi _{\uparrow}^{}(x)
\phi _{\downarrow}^{}(x) dx  \label{Ham-1}
\end{eqnarray}
describes 1D $\delta$-function interacting spin-$\frac{1}{2}$ Fermi gas of
$N$ fermions  with mass $m$ by periodic boundary conditions to a line of length $L$.   The field
operators $\phi_{\downarrow}$ and $\phi_{\uparrow}$ describe the
fermionic atoms in the states $|\uparrow \rangle$ and $\downarrow
\rangle$, respectively.    We use units of $\hbar =2m=1$ and denote 
coupling constant  $g_{\rm 1D}=\hbar^{2}c/m$ with $c=-2/a_{\rm 1D}$  where
$a_{1D}$ is the effective 1D scattering length \cite{Olshanii} $
a_{\rm 1D}=-\frac{a_{\perp}^2}{a_{\rm 3D}}+Aa_{\perp}$.
Here $a_{ \rm 3D}$ is  the 3D scattering length, $a_{\perp}=\sqrt{\hbar/(m\omega_{\perp})}$ is 
the transverse oscillator length, and $A\approx 1.0326$ is a numerical constant. 
 For repulsive fermions, $c>0$ and for attractive fermions, $c<0$.

 For an irreducible representation
$R_{\psi}=[2^{N_{\downarrow}},1^{N_{\uparrow}-N_{\downarrow}}]$ \cite{LGuan},
where  $N_{\uparrow}$ and $N_{\downarrow}$  are the numbers of fermions at the two  hyperfine levels $|\uparrow \rangle $ and $|\downarrow \rangle$  such that $N_{\uparrow}\geq N_{\downarrow}$.  The energy eigenspectrum is given in terms of the quasimomenta $\left\{k_i\right\}$  of the fermions via
$E=\frac{\hbar ^2}{2m}\sum_{j=1}^Nk_j^2$,
satisfying  the BA equations 
\begin{eqnarray}
&&\exp(\mathrm{i}k_{i}L)=\prod_{\alpha=1}^{M_1}\frac{k_{i}-\lambda_{\alpha}+\mathrm{i}c'}{k_{i}-\lambda_{\alpha}-\mathrm{i}c'},\label{BE-1}\\
&&\prod_{j=1}^{N}\frac{\lambda_{\alpha}-k_{j}+\mathrm{i}c'}{\lambda_{\alpha}-k_{j}-\mathrm{i}c'}=-\prod_{\beta=1}^{M_1}
\frac{\lambda_{\alpha}-\lambda_{\beta}+\mathrm{i}c}{\lambda_{\alpha}-\lambda_{\beta}-\mathrm{i}c},\label{BE-2}\\
&&i=1,2,\ldots,N, \qquad \alpha=1,2,\ldots,M_1 \nonumber
\end{eqnarray}
with the quantum number $M_1=N_{\downarrow}$ and a notation $c'=c/2$. The parameters $\left\{\lambda_{\alpha}\right\}$ are the
rapidities for the internal hyperfine spin degrees of freedom.

In thermodynamic limits, and for attractive regime, i.e. $c<0$,  quasimomenta $\left\{ k_i\right\} $ of the fermions with different spins  form two-body bound states,
i.e., $k_\alpha=\lambda_\alpha \pm   \mathrm{i} \frac{1}{2} c$, accompanied by the real spin parameter
 $\lambda_\alpha$ \cite{Yang-a,Takahashi-a}.  Here $\alpha =1,\ldots,M_1$.  The excess fermions have real quasimomenta $\left\{ k_j\right\} $ with $j=1,\ldots, N-2M_1$.  From these root patterns, the BA equations   (\ref{BE-1})  and (\ref{BE-2}) become 
\begin{eqnarray}
\exp(\mathrm{i}k_{i}L)&=&\prod_{\alpha=1}^{M_1}\frac{k_{i}-\lambda_{\alpha}+\mathrm{i}c'}{k_{i}-\lambda_{\alpha}-\mathrm{i}c'}, \label{BEa}\\
\exp(2\mathrm{i}\lambda_{\alpha }L)&=&\prod_{\ell =1}^{N-2M_1}\frac{\lambda_\alpha -k_\ell+\mathrm{i}c'}{\lambda_\alpha -k_\ell-\mathrm{i}c'}\prod_{\beta=1}^{M_1}\frac{\lambda_\alpha -\lambda_{\beta}+\mathrm{i}c}{\lambda_{\alpha}-\lambda_{\beta}-\mathrm{i}c},\label{BEb}\\
i&=&1,2,\ldots, N-2M_1,\qquad \beta=1,2,\ldots, M_1.\nonumber
\end{eqnarray}
In the above equations $\alpha \ne \beta$. 

\section{Polaron-Molecule crossover}

\subsection{Polaron-like state}
McGuire studied exact eigenvalue problem of $N-1$ Fermions of  the same spin and  one fermion of the opposite spin  in 1965  and 1966 \cite{McGuire}.  He calculated the energy shift caused by  this extra spin-down fermion.   The highly polarized Fermi system was   studied recently by Giraud and Combescot \cite{Giraud} in the context of Fermi polarons.  The polaron-like effect in the 1D Fermi-Hubbard model was studied by the variational anastz in \cite{Leskinen}.
 Here we prove that  a  single  spin-down fermion immersed into the a fully ploarized spin-up Fermi sea with weak attraction is likely to  behave like a polaron   in such  fermionic medium.  For the weak coupling  limit $L|c|\ll 1$,  we find that   either the spin-down fermion and a spin-up fermion from  the medium form a pair  $k_{\downarrow,\uparrow}=p\pm \mathrm{i} \beta$   or the spin-down fermion with a quasimomentum $k_{\downarrow}=p$ propagates in the medium. Whereas the rest are  $N-2$ real roots $\left\{k_i\right\}$ with $i=1,\ldots,N-2$. 
 
We consider  the quasimomenta of a pair $k_{\downarrow,\uparrow}=p\pm \mathrm{i} \beta$ and   $N-2$ real roots $\left\{k_i\right\}$ with $i=1,\ldots,N-2$.  From the discrete Bethe ansatz equations (\ref{BE-1}),  we have 
\begin{eqnarray}
e^{2\mathrm{i}pL} &=&\frac{k_{\downarrow}-p+\mathrm{i}c'}{k_{\downarrow}-p-\mathrm{i}c'}\frac{k_{\uparrow}-p+\mathrm{i}c'}{k_{\uparrow}-p-\mathrm{i}c'}\nonumber\\
e^{-2\beta L} &=&\frac{k_{\downarrow}-p+\mathrm{i}c'}{k_{\downarrow}-p-\mathrm{i}c'}\frac{k_{\uparrow}-p-\mathrm{i}c'}{k_{\uparrow}-p+\mathrm{i}c'}\nonumber\\
e^{\mathrm{i}k_{i}L}&=&\frac{k_{i}-p+\mathrm{i}c'}{k_{i}-p-\mathrm{i}c'}\label{BA-polaron}
\end{eqnarray}
with $i=1,\ldots,N-2$. From the second equation in (\ref{BA-polaron}),  we determine the imaginary part $\beta$ from 
\begin{equation}
\beta L = \tanh ^{-1}\frac{\beta |c|}{\beta^2 +c^2/4}.\label{beta}
\end{equation}
We see that for weak coupling limit $\beta \to \sqrt{ |c|/L}$  whereas for strong coupling limit $\beta \to |c|/2 $. From the   Bethe ansatz equations   (\ref{BE-2}), we obtain 
\begin{eqnarray}
&&\frac{p-k_{\downarrow}+\mathrm{i}c'}{p-k_{\downarrow}-\mathrm{i}c'}\frac{p-k_{\uparrow}+\mathrm{i}c'}{p-k_{\uparrow}-\mathrm{i}c'}\prod_{\ell=1}^{N-2}\frac{p-k_{\ell}+\mathrm{i}c'}{p-k_{\ell}-\mathrm{i}c'}=1.\label{BA-polaron2}
\end{eqnarray}
Using  the BA equations  (\ref{BA-polaron}) and (\ref{BA-polaron2}), we find  that in  weak coupling limit the roots satisfy the following polynomial equations 
\begin{eqnarray}
k_i &\approx& \frac{2n_i\pi}{L} -\frac{|c|}{L(k_i-p)} \label{polaron-k2}\\
p&\approx &\frac{2n_p\pi}{L} -\frac{|c|}{2L} \sum_{\ell=1}^{N-2} \frac{1}{(p-k_{\ell})}\label{polaron-p3}
\end{eqnarray}
with $ i=1,\ldots,N-2$. According to the Fermi statistics, here $n_i=\pm 1,\pm 2,\ldots , (N-2)/2$ and  $n_p$ is an integer. The ground state configuration corresponding to $n_p=0$. 
In the weak coupling limit, we have $\beta =  \sqrt{ |c|/L}$ so that   $p \gg \beta$.   Thus  we see that  the bound pair  is not essential in the limit $L|c|\ll 1$. The key feature of the model in this  limit is  that  the spin-down fermion receives a mean field from the fully-polarized Fermi sea.  We consider the case $n_p\ne 0$, i.e. $p\ne 0$ for excitations. 
From Eq. (\ref{polaron-k2}), we can calculate the energy of  the system with a single spin-down fermion
\begin{eqnarray}
E&\approx &-2\beta^2+ 2p^2+\sum_{i=1}^{N-2}k_i^2\nonumber\\
&\approx &-2\beta^2-\frac{2|c|}{L}(N-2)+p^2+\sum_{n=1}^{N_{\uparrow}/2}\frac{8(n \pi)^2}{L^2}\nonumber\\
&& -4|c|\sum_{i=1}^{\frac{N_{\uparrow}}{2}-1} \frac{p^2}{L(k_i^2-p^2)}. \label{polaron-e}
\end{eqnarray}
In the above equations, symmetrization of the $N-2$  quansimomenta in the medium was considered, i.e. the real roots associating with $N-2$ spin-up fermions can be symmetrized by $k_i\approx -k_j$ up to the order of $c$. This is  mainly because the quansimomenta of the spin-up fermions just  have  an order of $|c|$ deviation from the ones of the free spin-up fermions.    For $p=0$, the result (\ref{polaron-e}) coincides with  the ground state energy  of the Fermi gas with one spin-down fermion  given  by McGuire \cite{McGuire} (on page 125). If we consider the excitations for  the weakly bound  pair in the surrounding  fully-polarized Fermi sea, i.e. $p\ne 0$,  we can find an explicit relation of the energy depending on the total momentum of the system. Defining  total momentum of the system $q$, thus in  weak coupling limit,
we find a relation between $p$ and $q$ as 
\begin{equation}
p\approx q/\left(1-2|c|\sum_{i=1}^{\frac{N_{\uparrow}}{2}-1} \frac{1}{L(k_i^2-p^2)}\right).\label{p-q}
\end{equation}
Substituting (\ref{p-q}) into  (\ref{polaron-e}), the last term in (\ref{polaron-e}) is cancelled out.  Then we  obtain   an  energy shift 
\begin{eqnarray}
E(q,N,N_{\downarrow}=1) -E_{\uparrow}(N_{\uparrow},0)
&\approx& \epsilon_{p-b}+ \frac{\hbar^2q^2}{2m^{*}}
\label{polaron-aa}
\end{eqnarray}
that behaves like a quasi-particle polaron.  Where $E_{\uparrow}(N_{\uparrow},0)=\frac{\hbar^2}{2m}\frac{1}{3L^2}N_{\uparrow}^3\pi^2$ is the  kinetic energy of $N_{\uparrow}$ spin-up fermions. 
%$ \Delta \mu =-\frac{\hbar^2}{2m}n^2_{\uparrow}\pi^2$ is the variation of chemical potential.  
It is interesting to note that the   attractive mean field binding energy 
\begin{equation}
 \epsilon_{p-b}\approx \frac{\hbar^2}{2m}\left( -2\beta^2-\frac{2|c|}{L}(N-2)\right)\approx -\frac{\hbar^2}{m}n_{\uparrow}|c|
\end{equation}
  depends on  number of spin-up fermions and interaction strength \cite{Polaron-1}. 
 In the above equation (\ref{polaron-aa}),  the polaron-like state with an effective mass
$m^{*}\approx m(1+O(c^2))$ that is almost the same as the actual mass of the fermions in the limit $L|c|\ll 1$.  This is  consistent  with the result in \cite{McGuire,Giraud}.
The last term of  the equation (\ref{polaron-e2})  indicates that an  attractive interaction  always enhances the effective  mass of the polaron.  The addressed ``binding energy" can be rewritten as 
\begin{equation}
\epsilon_{p-b}=-\frac{6}{\pi^2}e_F|\gamma|,
\end{equation}
where the dimensionless interaction strength is defined $\gamma=c/n$. The fermi energy is $e_F=\frac{\hbar^2}{2m}\frac{1}{3}n^2\pi^2$. This  binding energy indicates  a mean field effect.  This is a 1D analog of Fermi polaron-like state resulted in from the weak attraction between the  impurity and the fully polarized Fermionic medium.  This polaron-like state  also exists   for weakly repulsive interaction, where the spin-down fermion  experiences a repulsive mean field energy shift. 

The mean field polaron-like state occurs for a few spin-down fermions immersed into a fully polarized Fermionic sea.
In thermodynamic limit and in weak coupling regime, i.e., $cL/N \sim 1$,  the ground state is the BCS-like  pairing  state with a pairing correlation length larger than the average interparticle  spacing. The
correlation function for the single particle Green's function decays
exponentially, i.e.,
$\langle\psi_{x,s}^{\dagger}\psi_{1,s}\rangle\rightarrow e^{-
x/\xi}$ with $\xi=v_F/\Delta$ and $s=\uparrow,\, \downarrow$,
whereas the singlet pair correlation function decays as a power of
distance, i.e.,
$\langle\psi_{x,\uparrow}^{\dagger}\psi_{x,\downarrow}^{\dagger}\psi_{1,\uparrow}\psi_{1,\downarrow}\rangle\rightarrow
x^{-\theta}$. Here $\Delta$ is the energy gap, and the critical
exponents $\xi$ and $\theta$ are both greater than zero. However,  once the external field exceeds the critical value, the Cooper pairs are destroyed. Thus both of these
correlation functions decay as a power of distance and the pairs
lose their dominance,  where, molecule and excess fermions form the polarized FFLO pairing-like phase.  In this phase 
the spacial oscillations of pairing correlation are   caused by an
imbalance in the densities of spin-up and spin-down fermions, i.e.,
$n_{\uparrow}-n_{\downarrow}$, which gives rise to a mismatch in
Fermi surfaces between both species of fermions.   In 1D,  the
pair and spin correlations with the spacial oscillation signature  are a consequence of  the  backscattering for bound pairs and unpaired fermions \cite{Lee-Guan}. 
In next section, we will see that  the polaronic signature is significantly different from the molecule state where single spin-down fermion and a spin-up fermion from the medium with a strong attraction form a tight bound molecule of two-atom. 

\subsection{The molecule state}

In order to catch the signature of molecule state,  we consider a spin-down fermion immersed into fully polarized spin-up medium with strong attraction, i.e. $ L|c|\gg  1$. We assume the bound pair  $k_{\downarrow,\uparrow}=p\pm \mathrm{i} \beta$ and   $N-2$ real roots $\left\{k_i\right\}$ with $i=1,\ldots,N-2$. From   (\ref{BE-1}) and  (\ref{BE-2}) with an odd number $N_{\uparrow}$, we find the real roots 
\begin{equation}
k_i\approx \frac{n_j\pi}{L}\left(1-\frac{4}{L|c|}\right)^{-1}-\frac{4P}{L|c|}\left(1-\frac{4}{L|c|}\right)^{-1},
\end{equation}
with $n_j=\pm 1, \pm 3,\ldots, \pm(N_{\uparrow}-1)$.
The pair excitations correspond to $p\ne 0$. Then we obtain the energy 
\begin{eqnarray}
E&=&-2\beta^2 +2p^2+\sum_{i=1}^{N-2}k_i^2\nonumber \\
&=& -2\beta^2 +2p^2 +E_{\uparrow}(N_{\uparrow},0)+\frac{16p^2(N_{\uparrow}-1)}{L^2c^2}.\label{e-m}
\end{eqnarray}
In the above equation, 
\begin{equation}
E_{\uparrow}(N_{\uparrow},0)=\frac{\hbar^2}{2m}\frac{1}{3L^2}N_{\uparrow}^3\pi^2 \left(1-\frac{4}{L|c|}\right)^{-2}
\end{equation}
 is the  kinetic energy of $N_{\uparrow}$ spin-up fermions with one spin-down fermion.  If we consider the  pair excitations with total momentum $q$, we find the relation between the $p$ and the total momentum of the system $q$, i.e.
 \begin{equation}
 p\approx q/\left[ 2\left( 1-\frac{2(N_{\uparrow}-2)}{L|c|} \right)\right].\label{p-q-m}
 \end{equation}
  Substituting (\ref{p-q-m}) into  (\ref{e-m}) , then we  obtain   an  energy shift 
\begin{eqnarray}
E(q,N,N_{\downarrow}=1) -E_{\uparrow}
&\approx& \varepsilon_b^t+ \frac{\hbar^2q^2}{2m^{*}}-\Delta\mu
\label{polaron-e2}
\end{eqnarray}
that behaves like a molecule with a binding energy 
\begin{equation}
\varepsilon_b^t=-2\beta ^2+ \frac{8\pi^2}{3|\gamma|}\approx -\frac{\hbar^2}{2m}\frac{c^2}{2} + \frac{8\pi^2}{3|\gamma|}
\end{equation}
  in the strong attractive regime $L|c|\gg 1$.  In the above equations $E_{\uparrow}=\frac{\hbar^2}{2m}\frac{1}{3L^2}N_{\uparrow}^3\pi^2$ is the  kinetic energy of $N_{\uparrow}$ spin-up fermions.  However, the effective mass of the molecule 
\begin{equation}
m^{*}\approx 2m\left(1-\frac{4(N_{\uparrow}-2)}{L|c|}\right)
\end{equation}
is almost twice the actual mass of the fermions.  In the above equations, the variation of the chemical potential $\Delta \mu=n_{\uparrow}^2\pi^2$. We see clearly that the system undergos a cross-over from a mean field polaron-like nature into  a mixture of excess fermions and a bosonic molecule as the attraction changes from a weak attraction  into a strong attraction.

\section{Ground state energy}

\subsection{Weak attraction }
 In order to see physical signature of the ground state energy at vanishing interaction strength, we first focus on  weakly attractive interaction in  which two  fermions with spin-up and 
spin-down states form a weakly  bound  pair. 
 In this regime, the weak bound  pair is 
not stable because the kinetic energy of the pair is larger than the binding energy.  In the limit $c\to 0^-$, the unpaired
fermions sit on two outer wings in the quasimomentum space
due to the Fermi statistics.  In this weak coupling  limit, i.e. $L|c|<<1$, the imaginary part of the pseudomomenta for
a BCS pair is proportional to $\sqrt{c/L}$.  Thus the bound state has
a small binding energy $\epsilon_{\rm b}={\hbar^2|c|}/{mL}$ that is 
 proportional to  $|c|$.  
For arbitrary polarization, the system is described by $M_1$ weakly bound pairs with $k_{\alpha}^{\rm p}\approx \lambda _{\alpha}\pm
\mathrm{i}\sqrt{c/L}$ and $N-2M_1$ unpaired fermions with real $k_i$.  Without losing generality, we assume $M_1$ is odd and $N$ is even. Substituting this root patterns into the BA equations (\ref{BE-1}) and  (\ref{BE-2}), we find the following equations to determine the  positive roots $\left\{\lambda_{\alpha}\right\}$ and the positive real quasimomenta $\left\{ k_j\right\}$ by 
\begin{eqnarray}
k_j&\approx & \frac{2n_j\pi}{L}+\frac{c}{Lk_j}+\frac{c}{L}\sum_{\alpha=1}^{\frac{1}{2}(M_1-1)}\left[\frac{2k_j}{k_j^2-\lambda_{\alpha}^2}\right],\label{BAd-1}\\
\lambda_{\alpha}&\approx  & \frac{2n_{\alpha}\pi}{L}+\frac{3c}{2L\lambda_{\alpha}}+\frac{c}{L}\sum_{
\beta =1}^{\frac{1}{2}(M_1-1)}\left[ \frac{2\lambda_{\alpha} }{\lambda_{\alpha}^2-\lambda_{\beta}^2}\right]\nonumber\\
&&+\frac{c}{2L}\sum_{j=1}^{\frac{1}{2}(N-2M_1)}\left[ \frac{2\lambda_{\alpha}}{\lambda_{\alpha}^2-k_j^2}\right],\label{BAd-2}
\end{eqnarray}
where $n_j=\frac{M_1+1}{2},\frac{M+3}{2},\ldots, \frac{N-M_1-1}{2}$,  and $n_{\alpha}=1,2,\ldots, \frac{M_1}{2}$. In the equation (\ref{BAd-2}), ${\alpha}=\beta$ is excluded.   By iteration, we obtain the ground state energy   for weakly attractive regime 
\begin{eqnarray}
E&=&-\frac{2M_1|c|}{L}+4\sum_{\alpha=1}^{\frac{1}{2}(M_1-1)}\lambda^2_{\alpha}+2\sum_{j=\frac{1}{2}(M_1+1)}^{\frac{1}{2}(N-M_1-1)}k_j^2\nonumber\\
&=&-\frac{2|c|(N-M_1)M_1}{L}+\frac{2\pi^2M_1(M_1^2-1)}{3L^2}\label{e-dis}\\
&&+\frac{\pi^2(N-2M_1)[N^2+M_1^2-M_1N-1]}{3L^2}+O(c^2).\nonumber
\end{eqnarray}
If we define linear density $n=N/L$ and the density of down-spin fermions $n_{\downarrow}=M_1/L$.  In thermaldynamic limit,  then the ground state energy per length (\ref{e-dis}) becomes 
\begin{eqnarray}
\frac{E}{L}=\frac{1}{3}n_{\uparrow}^3\pi^2+\frac{1}{3}n_{\downarrow}^3\pi^2+2cn_{\uparrow}n_{\downarrow}+O(c^2)\label{e-dis-w}
\end{eqnarray}
that agrees with the result given in \cite{BBGO}.
The  the ground state energy (\ref{e-dis}) is  also valid for weakly repulsive interaction, i.e., for $c>0$.  This gives a mean field effect for the 1D delta-function interacting Fermi gas in weak coupling regime.

\subsection{Strong attraction}

For strong attraction, i.e.  $L|c| \gg 1$,  (or say $c\gg k_F$), 
 the discrete BA equations   (\ref{BE-1}) and  (\ref{BE-2})  with the root patterns 
  $k_{\alpha}^{\rm p} =\lambda_{\alpha}\pm   \mathrm{i} \frac{1}{2} c$ for pairs and $k_j^{\rm u}$ for unpaired fermions
 can be linearized.  For even $N_{\downarrow}$,  we obtain the momenta of 
tight bound pairs and excess fermions \cite{BBGO}
\begin{eqnarray}
k_{\alpha} ^{(\rm p)} &\approx & \frac{n_{\alpha} \pi}{2L}\left(1+\frac{N_{\downarrow}}{Lc}+\frac{2(N-2N_{\downarrow})}{Lc}\right)^{-1} \pm \frac{1}{2} {\mathrm i} c,\\
 k_j^{(\rm u)}&\approx &\frac{n_j\pi}{L}\left(1+\frac{4N_{\downarrow}}{Lc}\right)^{-1} 
\end{eqnarray}
with integers 
$n_j=\pm 1, \pm 3,\ldots, \pm(N-2N_{\downarrow}-1)$ and  $n_{\alpha} =\pm 1,2, \ldots, \pm (N_{\downarrow}-1)$. 
In this scenario the bound states behave like
hard-core bosons due to Fermi statistics.
The per length ground state energy of  the model with strong
attraction and arbitrary polarization is given by 
\begin{equation}
\frac{E}{L}=E_0^u+E_0^b+n_{\downarrow} \varepsilon_b^t,
\end{equation}
where   $n_{\downarrow}=N_{\downarrow}/L$ and  the binding energy $\varepsilon_b^t=-\frac{c^2}{2}$ and the effective energy for unpaired fermions and pairs 
\begin{eqnarray}
E_0^u& \approx &\frac{(n-2n_{\downarrow})^3\pi^2}{3}\left[1+\frac{8n_{\downarrow}}{|c|}+\frac{48n_{\downarrow}^2}{c^2} \right],\label{E0u-dis-sa}\\
E_0^b& \approx& \frac{n_{\downarrow}^3\pi^2}{6}\left[1+\frac{2(2n_{\uparrow}-n_{\downarrow})}{|c|}  +\frac{3(2n_{\uparrow}-n_{\downarrow})^2}{c^2} \right].\label{E0b-dis-sa}
\end{eqnarray}
From the energies (\ref{E0u-dis-sa}) and (\ref{E0b-dis-sa}), we see that  the bound pairs have tails and the interfere with each other. But, it is impossible to separate the intermolecular forces from the interference between  molecules and  single fermions. 

\section{Relationship between the two sets of the Fredholm equations }
\label{FH}

 The fundamental physics of the model are determined by the set of transcendental equations  which can be transformed to the generalised Fredholm  equations in the thermodynamic limit.  This transformation  was found by Yang and Yang in series of papers on the study of spin $XXZ$ model  in 1966, see an  insightful  article  by Yang \cite{Yang-insight}. For repulsive interaction, the Bethe ansatz quasimomenta $\left\{ k_i\right\}$ are real, but all  $\left\{ \lambda_{\alpha }\right\}$ are real only for  the ground state. There are complex roots of $\lambda_{\alpha}$ called spin strings for excited states. In the thermodynamic limit, i.e., $L,N \to \infty$, $N/L$ is finite, the above Bethe ansatz equations (\ref{BE-1}) and (\ref{BE-2})  can be written as the generalized Fredholm equations 
\begin{eqnarray}
{r_1}(k)&=&\frac{1}{2\pi}+ \int_{-B_2}^{B_2}K_1(k-k'){r_2}(k')dk', \label{BE2-r1}\\
{r_2}(k)&=&\int_{-B_1}^{B_1}K_1(k-k'){r_1}(k')dk \nonumber\\
&&- \int_{-B_2}^{B_2}K_2(k-k'){r_2}(k') dk' ,\label{BE2-r2}
\end{eqnarray}
where the integration boundaries $B_1$, $B_2$   are determined by
\begin{equation}
n:\equiv N/L=\int_{-B_1}^{B_1}{r_1}(k)dk,  \qquad M_1/L=\int_{-B_2}^{B_2}{r_2}(k')dk'.\label{repulsive-d}
\end{equation} 
In the above equations,  we denote the function 
\begin{equation}
K_{m}(x)=\frac{1}{2\pi}\frac{mc}{(mc/2)^2+x^2}\label{a-r}
\end{equation}
with $c>0$ for repulsive regime and $c<0$ for attractive  regime.  The ground state energy per unit length is given by 
\begin{equation}
E=\int_{-B_1}^{B_1}k^2{r_1}(k) d k. \label{E3-r} 
\end{equation}
The functions  $r_m(k)$ denote  the Bethe ansatz root   distributions in parameter spaces, i.e. $r_1(k)$ stands for  quasimomenta distribution function, whereas $r_2(k)$  are the distribution functions for  rapidity parameter $\lambda$ in the BA equations  (\ref{BE-1}) and  (\ref{BE-2}). The ground state energy and full phase diagram can be obtained by solving analytically the Fredholm equations. 

For attractive regime, i.e. $c<0$,  the BA equations   (\ref{BEa})  and (\ref{BEb}) become 
\begin{eqnarray}
k_j L& =& 2\pi I_j + \sum_{l=1}^{M_1}
\theta (\frac{k_j-\lambda_l}{c'}),  \label{BA-d1} \\
2 \lambda_j L &=& 2 \pi J_j +
\sum_{l=1}^{N-2M_1}\theta(\frac{\lambda_j-k_l}{c'})+ \sum_{l=1}^{M_1}
\theta (\frac{\lambda_j-\lambda_l}{2 c'}),\label{BA-d2}\\
&&j=2M+1,\ldots, N,\qquad j=1,\ldots, M_1,\nonumber
\end{eqnarray}
where $\theta(x)=2 \arctan x$,  and $I_j=-(N-2M_1-1)/2,-(N-2M_1-3)/2,\ldots,(N-2M_1-1)/2$ and $J_j=-(M_1-1)/2,\ldots, (M_1-3)/2,(M_1-1)/2$.
 In thermodynamic limit,   we introduced the density of unpaired fermions $\rho_1(k)=dI_j(k)/Ldk$ and the density of pairs $\rho_2(k)=dJ_j(k)/Ldk$.  
They   satisfy the following Fredholm equations \cite{Yang-a,Takahashi-a}
\begin{eqnarray}
\rho_1(k)&=&\frac{1}{2\pi}+\int_{-Q_2}^{Q_2}K_1(k-k' )\rho_2(k')dk'  \label{Fermi2-a1}\\
\rho_2(k)&=&\frac{2}{2\pi}+\int_{-Q_1}^{Q_1}K_1(k-k')\rho_1(k')dk'\nonumber\\
&&+\int_{-Q_2}^{Q_2}K_2(k-k')\rho_2(k')dk'. \label{Fermi2-a2}
\end{eqnarray}
The linear densities  are defined by
\begin{eqnarray}
\frac{N}{L}&=& 2\int_{-Q_2}^{Q_2}\rho_2(k)dk+\int^{Q_1}_{-Q_1}\rho_1(k)dk,\nonumber \\
\frac{M_1}{L}&=&\int_{-Q_2}^{Q_2}\rho_2(k)dk.\label{density-a}
\end{eqnarray}
The ground state energy per length is given by
\begin{equation}
E=\int_{-Q_2}^{Q_2}\left(2k^2-c^2/2\right)\rho_2(k)dk+\int_{-Q_1}^{Q_1}k^2\rho_1(k)dk. \label{Fermi2-E-a}
\end{equation}
   We  will  investigate the relationship between  the two sets of the Fredholm equations for the Fermi gas with repulsive and attractive delta-function interactions.

In the light of Takahashi's unification of the ground state energy of the spin-1/2 weakly interacting Fermi gas  \cite{Takahashi-70},  we first examine the relationship between the Fredholm equations for 1D Fermions with repulsive and with attractive  delta-function interactions. It is convenient to use Yang's operator notations  \cite{Yang-a} for   the Fredholm equations. Here we denote the integral operator 
\begin{equation}
k_n:\equiv \langle k |k_n| k' \rangle =\frac{1}{2\pi}\frac{nc}{n^2c^2/4+(k-k')^2}
\end{equation}
which is a symmetric function. We define the projection operators $A_i$ and its dual projection operators $\bar{A_i}$
\begin{eqnarray}
\left\{  \begin{array}{ll}
\langle k |A_i{r_i} \rangle=0,&\,\,\,{\rm for}\,\,\,|k|>B_i,\\
\langle k |A_i{r_i} \rangle=r_i(k) ,&\,\,\,{\rm for}\,\,\,|k|\le B_i,
\end{array}\right.\\
\left\{  \begin{array}{ll}
\langle k |\bar{A}_i{r_i} \rangle=r_i(k),&\,\,\,{\rm for}\,\,\,|k|>B_i,\\
\langle k |\bar{A}_i{r_i} \rangle=0,&\,\,\,{\rm for}\,\,\,|k|\le B_i,
\end{array}\right.,
\end{eqnarray}
where $i=1,2$.  Similar notations  are  carried out for attractive interaction regime.  

The Fredholm equations (\ref{BE2-r1}) and (\ref{BE2-r2}) for repulsive interaction regime can be rewritten in terms of  these operators 
\begin{eqnarray}
\left( \begin{array}{l} r_1\\ r_2  \end{array}  \right)=\left( \begin{array}{l} \frac{1}{2\pi} \\ \frac{1}{2\pi} \end{array} \right)+\left( \begin{array}{ll}0&K_1\\ -K_1&0\end{array}\right)\left( \begin{array}{ll}\bar{A}_1&0\\ 0&A_2\end{array}\right)\left(\begin{array}{l} r_1\\ r_2 \end{array} \right),\label{Fermi2-ro1}
\end{eqnarray}
where the integration boundaries $B_i$ satisfy the following conditions
\begin{eqnarray}
 \frac{N^1}{L}&=&\frac{B_1}{\pi}- \frac{1}{\pi}\int \langle k |A_2r_2\rangle G_+(B_1,k)dk, \label{F2-r-n1}\\
\frac{N^2}{L} &=&  \frac{B_2}{\pi}- \frac{1}{\pi}\int \langle k|\bar{A}_1r_1\rangle  G_-(k,B_2) dk, \label{F2-r-n2}
 \end{eqnarray}
 for repulsive  regime. Here we denoted 
\begin{equation}
G_{\pm}(x,y)=\tan^{-1}\frac{c}{2(x-y)}\pm \tan^{-1}\frac{c}{2(x+y)}.
\end{equation}

For attractive regime the Fredholm equations are rewritten as
\begin{eqnarray}
\left( \begin{array}{l} \rho_1\\ \rho_2  \end{array}  \right)=\left( \begin{array}{l} \frac{1}{2\pi} \\ \frac{1}{2\pi} \end{array} \right)+\left( \begin{array}{ll}0&K_1\\ -K_1&0\end{array}\right)\left( \begin{array}{ll}\bar{A}_1&0\\ 0&A_2\end{array}\right)\left(\begin{array}{l} \rho_1\\ \rho_2 \end{array} \right), \label{Fermi2-rho1}
\end{eqnarray}
where $Q_1$ and $Q_2$ are determined by  
\begin{eqnarray}
 \frac{N^{1}}{L}&=&\frac{Q_1}{\pi}- \frac{1}{\pi}\int \langle k|A_2\rho_2\rangle  G_+(Q_1,k)  dk,  \label{F2-a-n1}  \\
\frac{N^2}{L} &=&  \frac{Q_2}{\pi}-\frac{1}{\pi}\int \langle k|\bar{A}_1{\rho_1}\rangle  G _-(k,Q_2) dk.  \label{F2-a-n2}
\end{eqnarray}
We prove that under  a   mapping 
\begin{eqnarray}
r_1(k)\leftarrow \rightarrow \rho_1(k),\qquad r_2(k)\leftarrow \rightarrow \rho_2(k), \label{mapping2}
\end{eqnarray}
 the  Fredholm equations  Eq. (\ref{Fermi2-ro1}) with (\ref{F2-r-n1}), (\ref{F2-r-n2})  for repulsive regime and   the Fredholm equations (\ref{Fermi2-rho1}) with   (\ref{F2-a-n1}), (\ref{F2-a-n2})  for attractive regime are identical.  In the above equations $c>0$ for repulsive interaction  regime and $c<0$ for attractive interaction regime  are implied. 
 
 Furthermore, in repulsive regime, the Fredholm equations (\ref{BE2-r1})  and (\ref{BE2-r2})  with  the Fermi boundaries conditions (\ref{F2-r-n1}) and (\ref{F2-r-n2}) exhibit  a symmetry 
\begin{eqnarray}
&&{r_1}(k) \leftarrow \rightarrow  {r_2}(k),\qquad A_1\leftarrow \rightarrow  \bar{A}_2, \nonumber\\
&& \bar{A}_1\leftarrow \rightarrow  {A}_2,\qquad c\leftarrow \rightarrow -c. \label{SY-r}
\end{eqnarray}
This symmetry  relates to  the spin-up and spin-down reversal symmetry of  the model. This transformation maps the eigenstates with $N_{\downarrow}$ down-spin atoms and $N_{\uparrow}$ up-spin atoms one-to-one onto the eigenstates with $N_{\downarrow}$ up-spin atoms and $N_{\uparrow}$ down-spin atoms for the gas. 
Similarly, in  attractive regime, the Fredholm equations (\ref{Fermi2-a1})  and (\ref{Fermi2-a2})  with the Fermi boundaries conditions (\ref{F2-a-n1}) and (\ref{F2-a-n2}) preserve the spin reversal symmetry 
\begin{eqnarray}
&&{\rho_1}(k) \leftarrow \rightarrow  {\rho_2}(k),\qquad A_1\leftarrow \rightarrow  \bar{A}_2, \nonumber\\
 &&\bar{A}_1\leftarrow \rightarrow  {A}_2,\qquad c\leftarrow \rightarrow -c.
 \label{SY-a}
\end{eqnarray}
As a consequence of the mapping (\ref{mapping2}),   the ground state energy smoothly connects at vanishing interaction strength.  To see this point, we need to  unify  the ground state energy for weakly repulsive and weakly attractive regimes. 
For weakly repulsive regime,  the ground state energy is given by 
\begin{eqnarray}
E&=&\int_{-B_1}^{B_1}k^2\langle k|A_1{r_1}\rangle dk=\frac{B_1^3}{3\pi} \\
&& +\int_{-B_2}^{B_2}\langle k'|A_2{r_2}\rangle \left[\int_{-B_1}^{-B_1}k^2\langle k| K_1|k' \rangle dk \right]dk'.\nonumber 
\end{eqnarray}
Substituting (\ref{Fermi2-ro1}) into the above equations, we obtain the ground state energy per length 
\begin{eqnarray}
E&=&\frac{B_1^2}{3\pi}+\frac{1}{2\pi}\int_{-B_2}^{B_2}H(k,B_1)dk \label{E-r1}\\
&& -\int_{-B_2}^{B_2}\left[ \int_{|k'|>B_1} \langle k|K_1|k'\rangle \langle k' |\bar{A}_1{r_1}\rangle dk'\right]H(k,B_1)dk,\nonumber
\end{eqnarray}
where 
\begin{eqnarray}
H(x,y)&=&\frac{1}{\pi}\left[(x^2-\frac{c^2}{4})\pi g_{y}(x)+yc \right. \nonumber \\
&&\left.  +\frac{1}{2}x c \ln \frac{4(x-y)^2+c^2}{4(x+y)^2+c^2}\right], \nonumber \\
g_y(x)&=&1-G_+(B_1,x).\nonumber
\end{eqnarray}

For attractive regime, the ground state energy  (\ref{Fermi2-E-a}) is rewritten as 
\begin{eqnarray}
\frac{E}{L} &=&\int_{-Q_2}^{Q_2}\left(2k^2-\frac{c^2}{2}\right)\langle k|A_2{\rho_2}\rangle dk \label{E-a-p}\\
&& +\int_{-Q_1}^{Q_1}k^2\left[\frac{1}{2\pi}-\int_{-Q_2}^{Q_2}\langle k|K_1| k' \rangle \langle k' |A_2 {\rho_2}\rangle dk' \right]dk.\nonumber
\end{eqnarray}
Substituting the Eq. (\ref{Fermi2-rho1}) into the above equation (\ref{E-a-p}), we obtain the ground state energy for  weakly attractive regime
\begin{eqnarray}
E&=& \frac{Q_1^2}{3\pi}+\frac{1}{2\pi}\int_{-Q_2}^{Q_2}H(k,Q_1)dk \label{E-a1}\\
&& -\int_{-Q_2}^{Q_2}\left[ \int_{|k'|>Q_1}\langle k|K_1|k' \rangle \langle k'|\bar{A}_1{\rho_1}\rangle  dk'\right]H(k,Q_1)dk.\nonumber 
\end{eqnarray}
We see that the ground state of the gas with a weakly repulsive interaction and a weakly attractive interaction  can be unified through   (\ref{E-r1}) and (\ref{E-a1}). 
The ground state energy  Eqs. (\ref{E-r1}) and Eqs. (\ref{E-a1}) can be calculated in a straight forward way. In weak coupling regime, it covers the result  (\ref{e-dis-w}), also see \cite{BBGO}.
The energy smoothly connects at vanishing interaction strength (but not analytically connects at  $c=0$).  Takahashi  \cite{Takahashi-70}  proved that the energy is infinitely differentiable at $c=0$ for a  real value of $c$.  But the two sets of the Fredholm equations turn to be divergent in the region $c\to \mathrm{i} 0$, see a discussion in (\cite{note2}).

\section{Conclusion}

In conclusion, we have studied  the polaron-molecule crossover in  the 1D spin-1/2 Fermi  gas with an attractive delta-function interaction. 
We have found that a spin-down fermion immersed into a   fully ploarized spin-up Fermi sea with weak attraction is  dressed  to form   a   Fermi polaron-like quasiparticle    in  the 1D   fermionic medium.  The spin-down fermion receives a mean field attraction  from the fully-polarized Fermi sea.  However, as the attraction grows,  the spin-down fermion binds with one  spin-up fermion  from the  fully-polarized medium to form a tightly bond molecule. We have presented the mean field binding energy and effective mass of the polaron in the weak attraction  limit and also presented   the binding energy and effective mass of the molecule in strong attraction  regime. 
The system undergos a cross-over from a mean field polaron-like nature into  a mixture of excess fermions and a bosonic molecule as the attraction changes from a weak attraction  into a strong attraction.  The asymptotic solutions of the discrete Bethe ansatz equations provides insight into understanding the mean field nature of the polaron-like state and the novel  pairing in the  interacting Fermi gas.   For both weak and strong coupling regimes, we have  obtained  the ground state energy from the Bethe ansatz roots of bound pairs and excess fermions, where they  have   the interfere with each other. Furthermore,  we  have proved that the two sets of the Fredholm  equations  for the 1D spin-1/2  Fermi gas  with  repulsive and attractive delta-function interactions are identical.  The result we obtained for weak and strong attractions opens to experimentally  study such mean field  nature of polaron-like state and molecule signature  in  1D trapped cold atoms.  Current experiment is capable of catching  the polaronic signature  of 1D interacting quantum gases of cold atoms. This can be  possibly achieved by using a species selective dipole potential, see a recent experiment on the polaronic dynamics of the 1D Bose gas \cite{Catani}. 

{\bf Acknowledgment.} This work is supported by the Australian Research Council. Author  thank Prof.  Tin-Lun Ho, Prof. Zhong-Qi Ma and Prof. Chen-Ning Yang for helpful discussions and encouragements. He also acknowledges the Zhong-Shan University  for their kind hospitality.

\end{document}